# Crystal structure of the new $FeSe_{1-x}$ superconductor†

**Serena Margadonna,*[a] Yasuhiro Takabayashi,[b] Martin T. McDonald,[b] Karolina Kasperkiewicz,[a] Yoshikazu Mizuguchi,[c] Yoshihiko Takano,[c] Andrew N. Fitch,[d] Emmanuelle Suard[e] and Kosmas Prassides*[b]**

**The newly discovered superconductor $FeSe_{1-x}$ ($x$~0.08, $T_c^{onset}$~13.5 K at ambient pressure rising to 27 K at 1.48 GPa) exhibits a structural phase transition from tetragonal to orthorhombic below 70 K at ambient pressure – the crystal structure in the superconducting state shows remarkable similarities to that of the $REFeAsO_{1-x}F_x$ (RE = rare earth) superconductors.**

Superconductivity at the surprisingly high temperature of 55 K has been recently reported in fluorine-doped rare-earth iron oxyarsenides, $REFeAsO_{1-x}F_x$.[1] The magnitude of $T_c$ and the apparent similarities with the high-$T_c$ cuprate superconductors – layered structural motifs of the conducting FeAs slabs and proximity to antiferromagnetic (AFM) and structural instabilities – have made these systems an intensely studied research field. The parent REFeAsO materials possess a simple tetragonal crystal structure (ZrCuSiAs-type, space group *P*4/*nmm*) comprising layers of edge-sharing $FeAs_4$ tetrahedra interleaved with REF layers. On cooling, a structural phase transition to orthorhombic crystal symmetry (space group *Cmma*), accompanied by the development of long range AFM order occurs.[2] The magnetic instability is suppressed upon fluorine-doping before the onset of superconductivity,[3] while orthorhombic symmetry survives well within the superconducting compositions.[4] Soon thereafter it was discovered that superconductivity can be induced by oxygen vacancies in the $REFeAsO_{1-\delta}$ systems.[5] The family of FeAs-based superconductors has now further expanded to include the related alkali-doped superconductors, $A_{1-x}A'_xFe_2As_2$ (A = alkaline earth, A′ = alkali metal)[6] as well as the ternary LiFeAs phase.[7]

These discoveries have catalysed the search for superconducting compositions in related materials in which two-dimensional FeQ (Q = non-metal ions) slabs are also present. This search has now led to the report that superconductivity at ~ 8 K occurs in the simple binary α-$FeSe_{1-x}$ phase field for $x$ = 0.12 and 0.18.[8] Subsequent work has revealed resistivity onsets for the superconducting transition at temperatures as high as 13.5 K at ambient pressure.[9] Quite remarkably, $T_c$ is extremely sensitive to applied external pressure and rises rapidly at a rate of 9.1 K $GPa^{-1}$, reaching a value of 27 K at 1.48 GPa.[9] α-FeSe crystallises at room temperature with the tetragonal PbO-type structure (Fig. 1) and

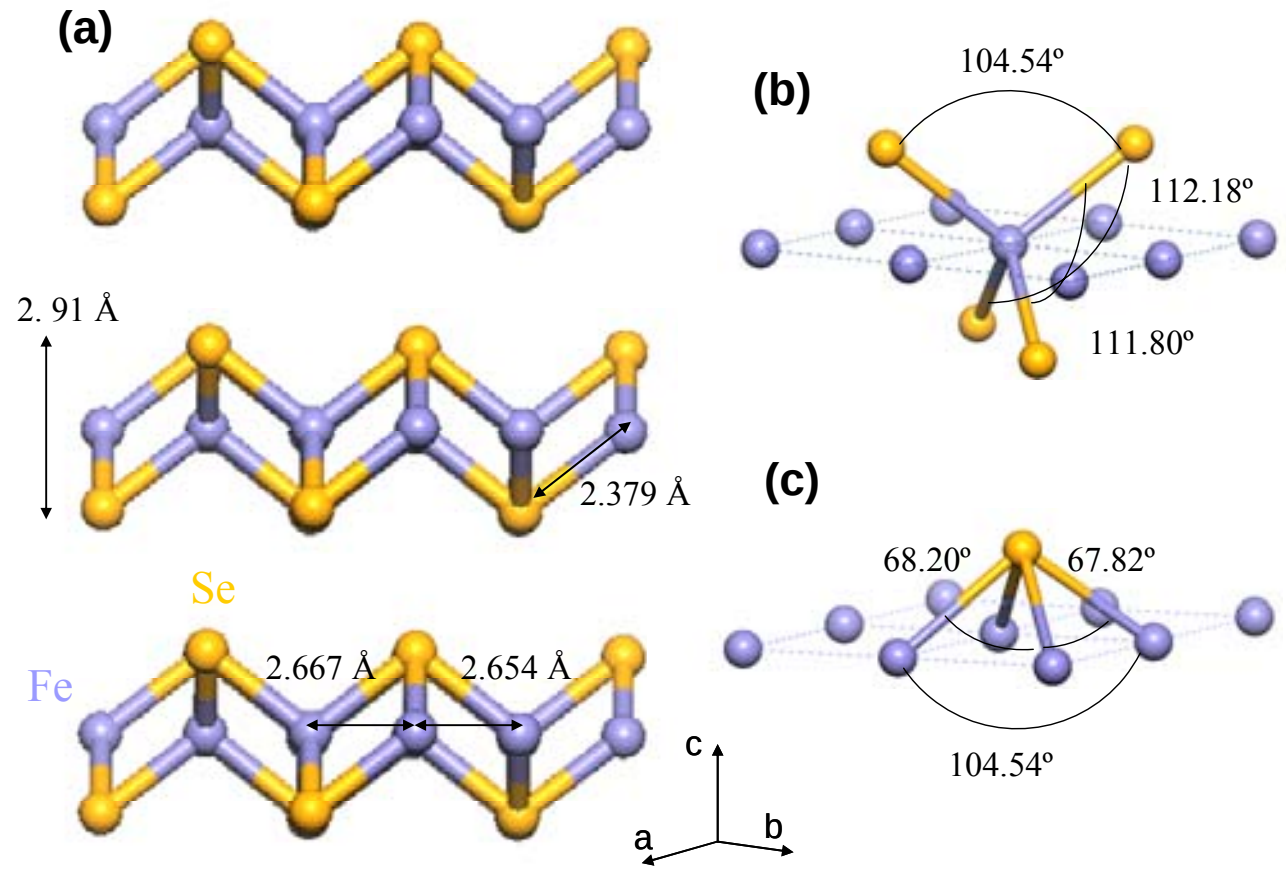

**Fig. 1 (a)** Schematic diagram of the low-temperature orthorhombic crystal structure of α-$FeSe_{0.92}$. Fe and Se ions are depicted as blue and yellow spheres, respectively. **(b)** and **(c)** Geometry of the $FeSe_4$ tetrahedra and the $SeFe_4$ pyramids with the three distinct Se-Fe-Se and Fe-Se-Fe bond angles indicated.

comprises stacks of edge-sharing $FeSe_4$ tetrahedra – the FeSe packing motif is essentially identical to that of the FeAs layers in the iron oxyarsenides. On cooling, it was reported that a structural phase transition occurs in the vicinity of 105 K to an unidentified low-temperature structure with possible triclinic symmetry.[8] At present, little else is known about the electronic properties of the superconducting α-$FeSe_{1-x}$ phases but it is reasonable to assume that Se non-stoichiometry is accompanied by the introduction of charge carriers in the essentially 2D FeSe layers. At this stage, precise structural information on the superconducting phase of α-$FeSe_{1-x}$ is needed and this should be the cornerstone for all subsequent understanding of the origin of superconductivity in this intriguing system.

Here we report the structural determination of the α-$FeSe_{1-x}$ ($x$ ~ 0.08) binary superconductor by high-resolution synchrotron X-ray and neutron powder diffraction at temperatures between 5 and 295 K. We find that below $T_s$ = 70 K, the crystal structure becomes metrically orthorhombic (space group *Cmma*), displaying an identical distortion of the FeSe slabs to that oberved for the FeAs layers in the iron oxyarsenide family. The structural transition coincides with the temperature at which the temperature-dependent resistivity, ρ shows an anomaly.

[a] *School of Chemistry, University of Edinburgh, Edinburgh, UK EH9 3JJ. E-mail: serena.margadonna@ed.ac.uk*
[b] *Department of Chemistry, University of Durham, Durham, UK DH1 3LE. E-mail: K.Prassides@durham.ac.uk*
[c] *National Institute for Materials Science, 1-2-1 Sengen, Tsukuba 305-0047, Japan.*
[d] *European Synchrotron Radiation Facility, 38043 Grenoble, France.*
[e] *Institut Laue Langevin, 38042 Grenoble, France.*
† Electronic Supplementary Information (ESI) available: results of the structural refinements. See http://dx.doi.org/10.1039/b000000x/

Powder samples of $FeSe_{1-x}$ were prepared by reaction of stoichiometric quantities of high-purity Fe (>99.9%) and Se (99.999%) contained in an evacuated quartz tube at 680°C for 12 hours. Following regrinding, these were pressed into pellets and heated in vacuo at 680°C for an additional period of 12-30 hours.[9] SQUID measurements (ZFC/FC protocols, 10 Oe) reveal the onset of bulk superconductivity below ~8 K. Resistivity measurements by the four-probe method show the onset of superconductivity at 13.5 K with the zero-resistance state obtained at ~7.5 K. In addition, two small humps in ρ(T) are evident near 200 and 50 K.[9] High-resolution synchrotron X-ray diffraction experiments were carried out on the ID31 beamline at the European Synchrotron Radiation Facility (ESRF), France. The samples were sealed in 0.7-mm diameter thin-wall glass capillaries and diffraction profiles (λ = 0.40301 Å) were collected at various temperatures between 5 and 295 K. The data were binned in the 2θ range 1-40° to a step of 0.002°. Higher statistics diffraction profiles were also recorded at 5, 100, 200 and 295 K over a longer angular range (2θ = 1° to 50°). Complementary neutron powder diffraction data were collected with the high-resolution diffractometer D2b (λ= 1.5944 Å) at the Institut Laue Langevin (ILL, Grenoble, France). The sample (149 mg), from the same batch used for the synchrotron X-ray diffraction measurements, was loaded in a cylindrical vanadium can (diameter = 5 mm) and then placed in a standard ILL "orange" liquid helium cryostat. The instrument was operated in its high-flux mode and the data were collected in the angular range, 2θ = 0°-164.5° in steps of 0.05°. Full diffraction profiles were measured with counting times of 8 hours at 5 and 295 K. The raw data were merged and normalised to standard vanadium runs using local ILL programmes. Analysis of the diffraction data was performed with the GSAS suite of Rietveld analysis programmes.

Inspection of the synchrotron X-ray and neutron diffraction profiles of α-$FeSe_{1-x}$ at ambient temperature confirmed the primitive tetragonal (T) unit cell ($a$ = 3.77376(2) Å, $c$ = 5.52482(5) Å; space group $P4/nmm$, $R_{wp}$ = 7.87% (SXRPD) and 5.80% (NPD), $\chi^2$ = 2.743, Fig. 1S, Table 1S). The Se content refines to a value of 0.91(9). Additional peaks were also evident in the profiles and these could be accounted for by the presence of a minority hexagonal β-$FeSe_{1-x}$ phase (space group $P6_3/mmc$, 22% fraction as revealed by the results of the combined Rietveld refinements). Based on the evolution of the diffraction profiles down to 90 K the refined structure remained strictly tetragonal with both lattice constants, $a$ and $c$ decreasing smoothly (Fig. 2). The rate of contraction at 13.5 and 30.0 ppm $K^{-1}$ for the $a$ and $c$ lattice constants, respectively is considerably anisotropic and leads to a gradual decrease of the ($c/a$) ratio with decreasing temperature. However, the lattice response to further decrease in temperature below 90 K is dramatically different – the $hkl$ ($h$, $k \neq 0$) reflections in the diffraction profile first develop a characteristic broadening, followed by a clear splitting at $T_s$ ~ 70 K (Fig. 3) signifying a lowering in symmetry of the high-temperature tetragonal structure and the onset of a structural transition. The magnitude of the splitting increases monotonically as the sample is cooled further down to 5 K.

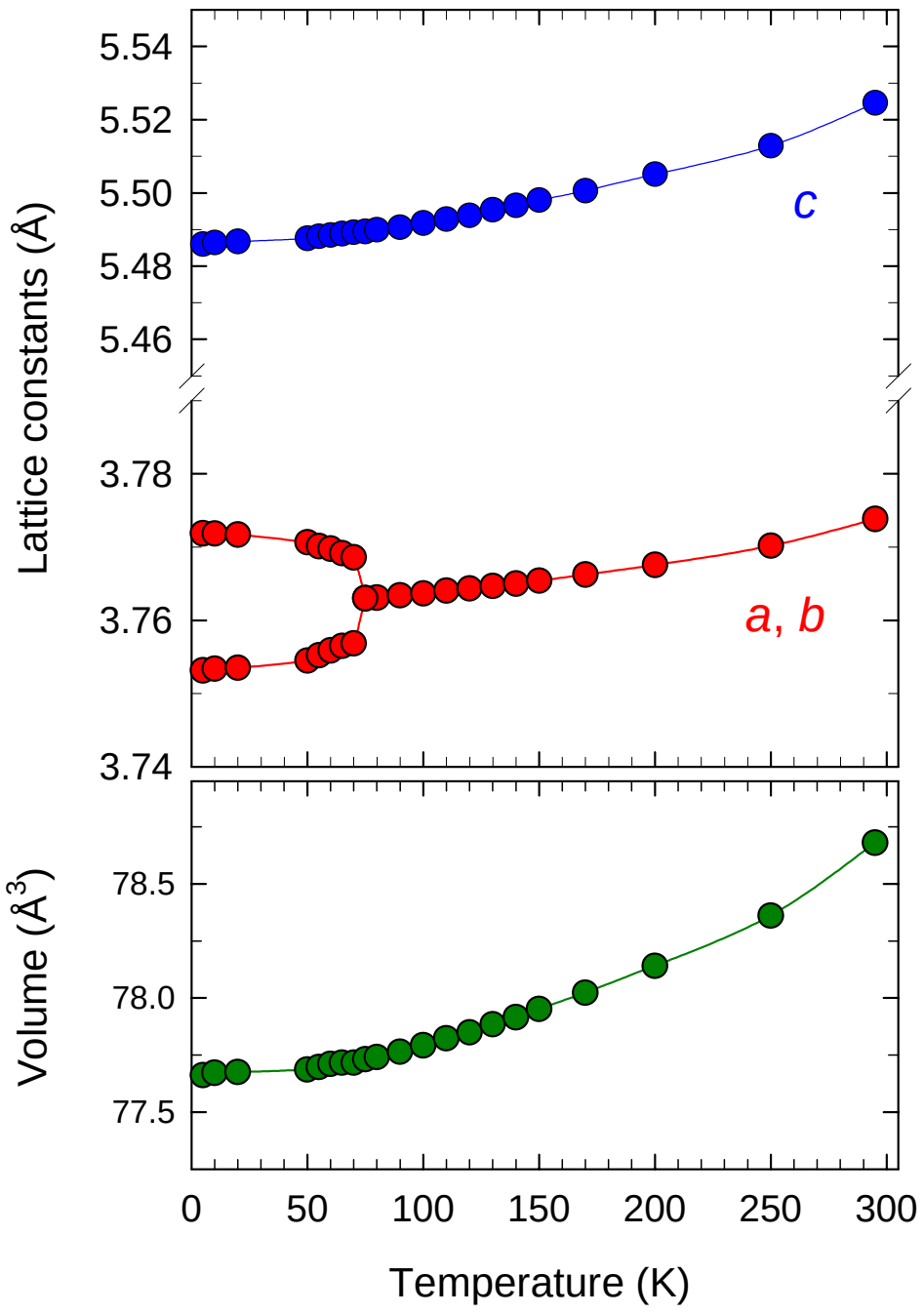

**Fig. 2** Temperature evolution of the lattice constants (top) and the unit cell volume, $V$ (bottom) in α-$FeSe_{0.92(1)}$. The $a$ and $b$ lattice constants are divided by √2 at temperatures below the tetragonal-to-orthorhombic phase transition.

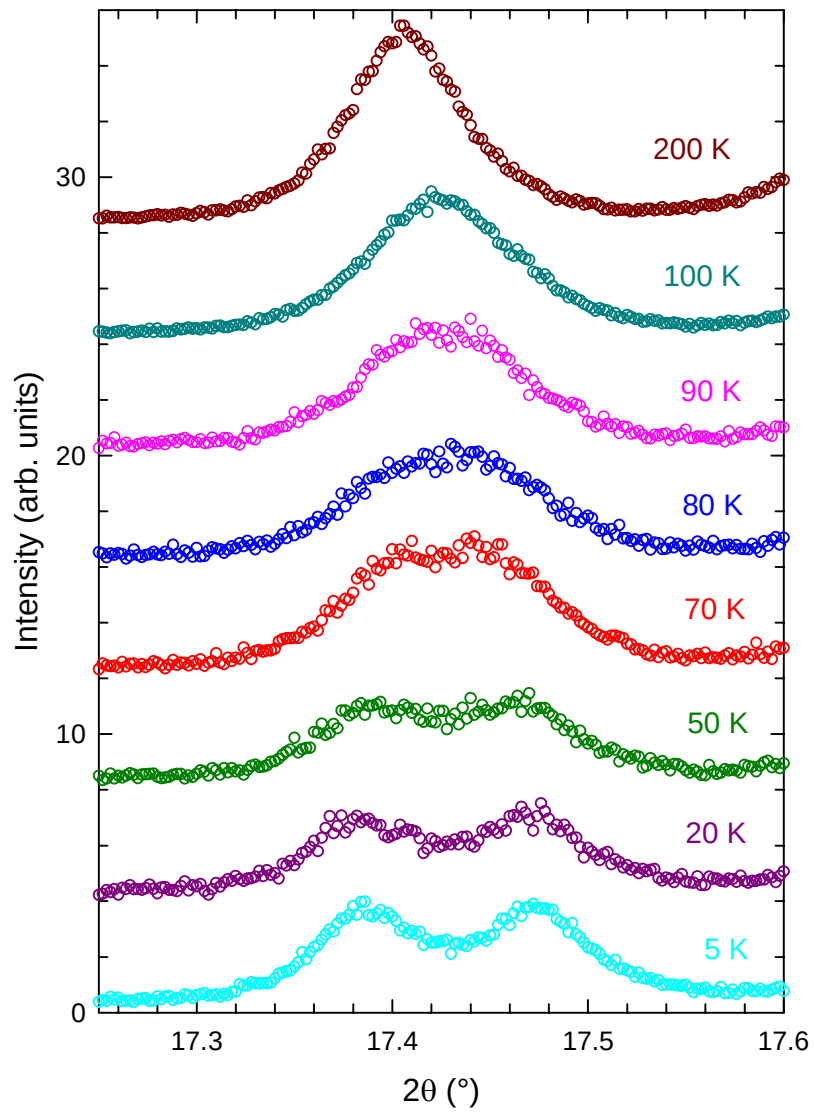

**Fig. 3** Selected region of the synchrotron X-ray diffraction profile of α-$FeSe_{0.92(1)}$ showing the temperature evolution of the $(220)_T$ Bragg reflection which on cooling splits into a doublet $[(040)_O, (400)_O]$ (λ = 0.40301 Å).

Combined Rietveld refinements of the synchrotron X-ray and neutron powder diffraction data of α-$FeSe_{1-x}$ at 5 K were performed successfully with the same orthorhombic (O) superstructure model (space group $Cmma$, $b > a \sim \sqrt{2}a_T$, $c \sim c_T$, where $a_T$ and $c_T$ are the lattice constants of the high-temperature tetragonal unit cell) employed for the low-temperature phases of the $REFeAsO_{1-x}F_x$ family.[2,4] At 5 K, the Se content refines to 0.92(1) and the refined lattice constants are $a$ = 5.30781(5) Å, $b$ = 5.33423(5) Å and $c$ = 5.48600(5) Å ($R_{wp}$ = 8.34% (SXRPD) and 3.77% (NPD), $\chi^2$ = 2.750, Fig. 4, Table 1S). The temperature evolution of the lattice constants and of the unit cell volume is shown in Fig. 2. The structural phase transition below 70 K is accompanied by the development of orthorhombic strain in the $ab$ basal plane – this increases continuously on cooling, approaching a value of $s = (b-a)/(b+a) = 2.5\times10^{-3}$ at 5 K. However, the structural transition has no influence on the magnitude of either the interlayer *c* lattice constant or the unit cell volume, *V* as both decrease smoothly on cooling across the tetragonal-orthorhombic phase boundary. The influence of the T→O phase transition on the local geometry of the $FeSe_4$ tetrahedra in the SeFeSe slabs is only quite subtle (Fig. 1) – while at high T, there is a single Fe-Fe distance (at 295 K: 2.668 Å), below $T_s$, there are now two symmetry inequivalent Fe-Fe distances (at 5 K: 2.667 and 2.654 Å). Similarly, one set of the Fe-Se-Fe angles remains unaltered (at 295 K: 104.3°, at 5 K: 104.5°), while the second set (at 295 K: 112.1°) splits into two (at 5 K: 111.8° and 112.2°). On the other hand, there is a single Fe-Se bond distance both in the tetragonal and orthorhombic phases (at 295 K: 2.389 Å, at 5 K: 2.379 Å).

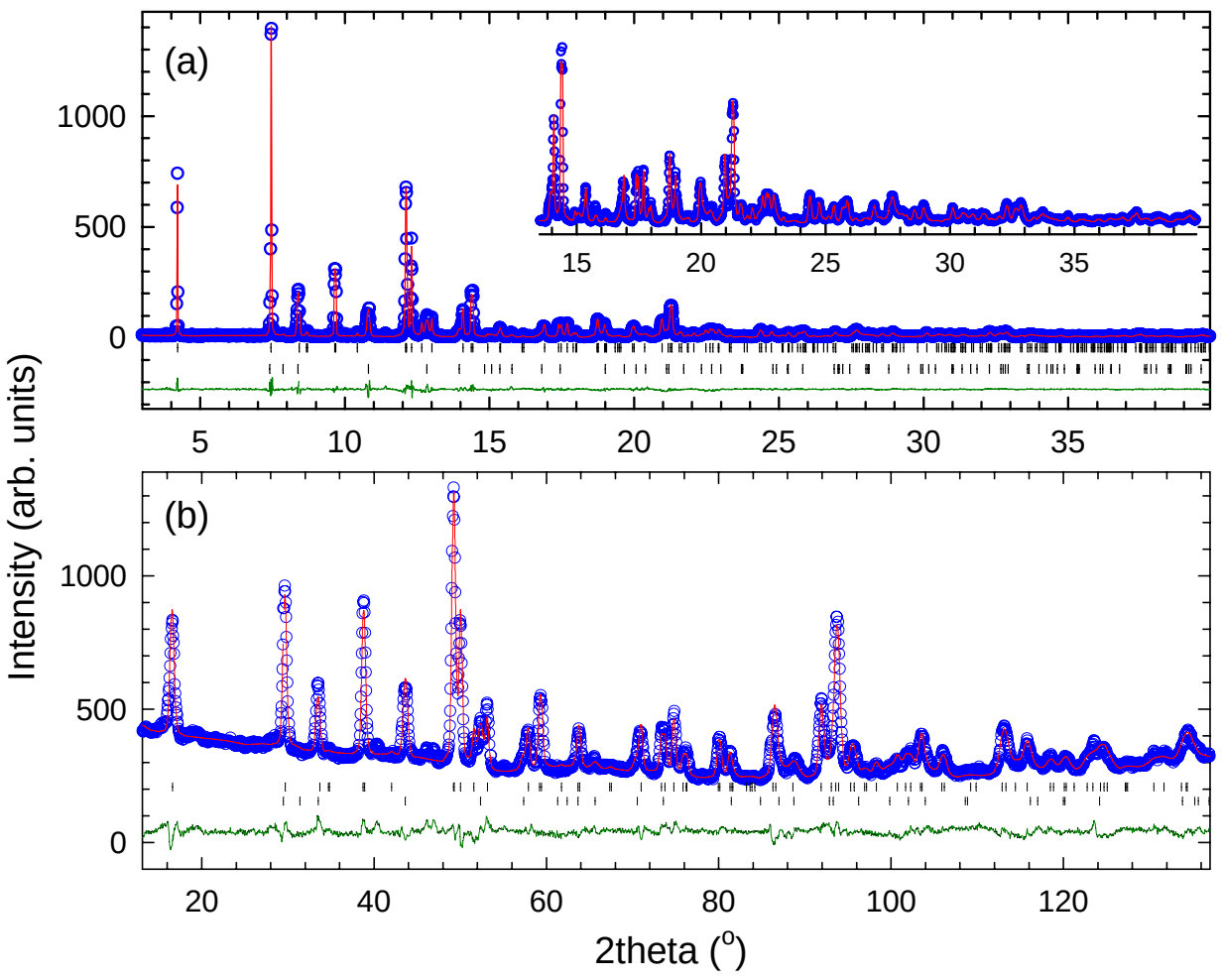

**Fig. 4** Final observed (circles) and calculated (solid lines) **(a)** synchrotron X-ray (λ = 0.40301 Å) and **(b)** neutron (λ = 1.5944 Å) powder diffraction profiles for the α-$FeSe_{0.92(1)}$ sample (78% fraction) at 5 K. The lower solid lines show the difference profiles and the tick marks show the reflection positions of the α- (top) and β-$FeSe_{1-x}$ (bottom) phases. The inset in **(a)** shows an expanded view of the diffraction profile at high Bragg angles.

An important point arising from the present structural refinements is that the crystal structure of the superconducting state at ambient pressure in α-$FeSe_{0.92}$ is metrically orthorhombic. Moreover, there is a signature of the structural transformation in the electronic properties as $T_s$ coincides with the observation of a hump in the temperature dependence of ρ(*T*). This is reminiscent of the structural behaviour in the $SmFeAsO_{1-x}F_x$ oxyarsenide family in which an isostructural orthorhombically distorted structure is also adopted for the superconducting compositions with F doping levels $x \leq 0.12$, while the T→O phase transition is only suppressed at doping levels, $x \geq 0.15$.[4] It is quite likely that an analogous suppression may occur for higher levels of Se non-stoichiometry in the present system. Certainly it is intriguing to conjecture that application of pressure may also lead to suppression of the low-temperature T→O phase transition, thereby accounting for the remarkable rapid enhancement of $T_c$ upon pressurisation.

Fig. 1 shows a schematic view of the orthorhombic structure of α-$FeSe_{0.92}$. It comprises SeFeSe slabs of ~2.91 Å thickness made of edge-sharing distorted $FeSe_4$ tetrahedra and separated from each other by an interlayer distance of ~2.58 Å. The $SmFeAsO_{1-x}F_x$ superconductors adopt a similar packing geometry made of AsFeAs slabs of ~2.7 Å thickness interleaved with Sm(F,O) layers which are absent in α-$FeSe_{0.92}$. The absence of the charge reservoir layers leads to a drastically reduced *c* lattice constant of ~5.5 Å for α-$FeSe_{0.92}$ (*cf.* *c* ~ 8.5 Å for $SmFeAsO_{1-x}F_x$). The basal plane lattice constants of α-$FeSe_{0.92}$ are also somewhat smaller than those in $SmFeAsO_{1-x}F_x$ reflecting the shorter Fe-Se bond lengths (2.38 Å vs 2.40 Å for the Fe-As bond lengths).

Finally, it has been argued before for the iron oxypnictide superconductors that the geometry of the $AsFe_4$ pyramidal units (Fig. 1) sensitively controls both the Fe near- and next-near-neighbour exchange interactions[10] and the width of the electronic conduction band.[11,12] As a result, the superconducting transition onset, $T_c$ of $REFeAsO_{1-x}F_x$ increases as the Fe-As-Fe angles become progressively smaller and the rare earth series is tranversed with the $RE^{3+}$ ionic radius decreasing. At present, there is no information available on the electronic structure of the iron selenide superconductors and the corresponding influence of the local geometry on the conduction bandwidth. Nonetheless we note that for α-$FeSe_{1-x}$, the two small Fe-Se-Fe angles at ~68° (Fig. 1c) are comparable to those in $SmFeAsO_{1-x}F_x$ (~71°), while the larger Fe-Se-Fe angle of the $SeFe_4$ pyramids at ~105° is considerably smaller that those of the iron oxyarsenides (>111°).

In conclusion, we have found that the α-$FeSe_{0.92}$ superconductor adopts at low temperatures an orthorhombic superstructure of the high-temperature tetragonal PbO-type structure. The structural transition at ~70 K is evidenced in the electronic properties via a hump in the temperature dependence of the resistivity. Given the very large positive pressure coefficient of $T_c$ (~9 K/GPa) in this material, it will be intriguing to follow the structural behaviour and the response of the orthorhombic distortion as a function of applied external pressure.

We thank the ESRF and the ILL for synchrotron X-ray and neutron beamtime, respectively.

## Notes and references